\documentclass[aps,prc,showpacs,twocolumn,amssymb,floatfix]{revtex4}
\usepackage{graphicx}
\usepackage{color}

\newcommand{\dd}  { {\textrm d}}
\newcommand{\dAu} {\mbox{$ dAu $}}
\newcommand{\AuAu} {\mbox{$ AuAu $}}
\newcommand{\pperp}{\mbox{$p_T$}}
\newcommand{\piz}{\mbox{$\pi^0$}}
\newcommand{\RdAu}{\mbox{$R_{dAu}$}}
\newcommand{\RdAupt}{\mbox{$R_{dAu}(\pperp)$}}

\begin{document}
\title{EMC effect and jet energy loss in relativistic deuteron-nucleus collisions}
\author{Brian A. Cole}
\affiliation{Nevis Laboratory, Columbia University,
New York, NY, USA}
\author{Gergely G\'abor Barnaf\"oldi, P\'eter L\'evai}
\affiliation{RMKI Research Institute for Particle and Nuclear Physics, \\
P.O. Box 49, Budapest 1525, Hungary}
\author{G\'abor Papp}
\affiliation{Department of Theoretical Physics, E\"otv\"os University,\\
P\'azm\'any P. 1/A, Budapest 1117, Hungary}
\author{George Fai}
\affiliation{Department of Physics, Kent State University, Kent, OH, USA}

\vspace*{.5cm}

\date{5 February 2007}

\begin{abstract}

We investigate the influence of modified nuclear parton
distribution functions (PDFs) on high-\pperp\ hadron production
at RHIC and LHC energies using a pQCD-improved parton model.
For application at RHIC, we focus on the possible contribution of the
EMC modification of the nuclear PDFs in the $x \gtrsim 0.3$ region to
the observed suppression of \piz\ production at $\pperp \gtrsim 10$~GeV/c
in \dAu\ collisions. We study three different parameterizations of the
nuclear PDF modifications and find that they give consistent results
for \RdAupt\ for neutral pions in the region $10$~GeV/c 
$\lesssim \pperp \lesssim 20$~GeV/c.  We
find that the EMC suppression of the parton distributions in the $Au$
nucleus does not strongly influence the \RdAu\ for \piz\ in the 
\pperp\ region where the suppression is observed. Using the HKN
parameterization, we evaluate systematic errors in the theoretical
\RdAupt\ resulting from uncertainties in the nuclear PDFs.  The
measured nuclear modification factor is inconsistent with the 
pQCD model result for 
$\pperp \gtrsim 10$~GeV/c even when the systematic uncertainties in the 
nuclear PDFs are accounted for. The inclusion of a small final-state 
energy loss can reduce the discrepancy with the data, but 
we cannot perfectly reproduce the \pperp\ dependence of the 
measured \RdAupt. 
For the LHC, we find that shadowing of the nuclear PDFs produces a large 
suppression in the yield of hadrons with $\pperp \lesssim  100$~GeV/c in 
$p(d)A$ collisions.  

\end{abstract}


\maketitle
\section{Introduction}

Deuteron-gold (\dAu) collisions have been extensively studied at the 
Relativistic Heavy Ion Collider (RHIC), both for their intrinsic 
interest and as a control experiment~\cite{PHENIXdAu,PHENIXdAu05,STARdAu} 
to judge the suppression seen in central gold-gold (\AuAu ) collisions
at sufficiently high transverse momenta ($p_T$)~\cite{PHENIXAuAu,STARAuAu}.
Unexpectedly, not only \AuAu\ data, but recent extended $p_T$ coverage 
\dAu\ data also display a suppressed nuclear modification factor
in central collisions\cite{PHENIXdAu06}.
This motivates a study of possible mechanisms that may result in a 
nuclear modification factor smaller than unity at sufficiently high 
transverse momenta ($p_T \gtrsim 6$ GeV/c) in central \dAu\ collisions. 
 
The nuclear modification factor $R_{dAu}$ compares the 
spectra of produced particles in \dAu\ collisions to a hypothetical scenario 
in which the nuclear collisions are assumed to be a superposition of the 
appropriate number of nucleon-nucleon collisions.
In the transverse momentum window $2$~GeV/c $\lesssim p_T \lesssim  6$~GeV/c, 
the nuclear modification factors are 
dominated by the Cronin peak~\cite{Cron75,Antr79}. Several physical 
pictures of this enhancement have been 
proposed~\cite{CronGyL98,CronKopel02,CronDima03,CronAcc04,CronBlaiz04,CronHwa}. 
One family of models~\cite{Wong98,Wang01,PLF00,Yi02,Papp02,BGG04} 
advances an explanation of the Cronin effect in terms of the interplay 
between nuclear 
shadowing~\cite{emc95,Shad_EKS,Shad_HIJ,Shadxnw_uj,Shad_HKM,Shad_HKN} and 
the multiple scattering of particles propagating in the 
strongly-interacting medium 
(multiscattering)~\cite{Wang01,PLF00,Yi02,Papp02}. Using the HIJING 
shadowing prescription~\cite{Shad_HIJ,Shadxnw_uj}, our model gave a 
reasonable description of the Cronin effect in central collisions at 
midrapidity~\cite{Levai0306,Levai0611}. At the same time, we obtained 
nuclear modification factors close to unity at high transverse momenta 
($6$~GeV/c $\lesssim p_T \lesssim  20$~GeV/c). 

In the high-$p_T$ region
multiscattering no longer affects $R_{dAu}$. It is then natural to ask if 
nuclear shadowing can explain suppression effects at high $p_T$. 
In particular, since at $\sqrt{s_{NN}} = 200$~$A$GeV we are in the EMC region of 
the shadowing function for $6$~GeV/c $\lesssim p_T \lesssim  20$~GeV/c, we ask 
if the EMC effect~\cite{emc95} plays a role in understanding these experiments.

In this paper we first investigate \dAu\ collisions at the highest RHIC 
energies and the role of the EMC effect 
at transverse momenta up to $50-70$ GeV/c. We then examine whether recent
shadowing parameterizations incorporating theoretical uncertainties for
the first time\cite{Shad_HKM,Shad_HKN} can account for the experimental information. 
We also display calculational results with a modest energy loss using energy-loss
parameters applied earlier to \AuAu\ data.
Finally we extend our considerations to the energy range of
the Large Hadron Collider (LHC).   


\section{The EMC effect in deuteron-gold collisions}
\label{sec:1}

\begin{figure}
\centerline{%
\rotatebox{0}{\includegraphics[width=\columnwidth,height=7.0truecm]%
   {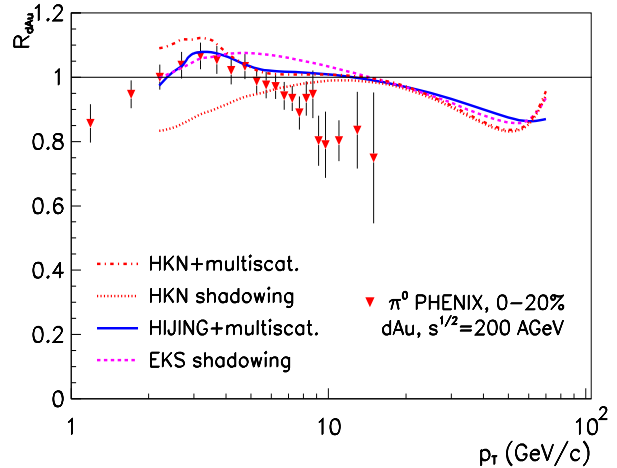}}
}
\caption{(Color online)
The nuclear modification factor, $R_{dAu}$ in central ($0-20 \ \%$) \dAu\ 
collisions for $\pi^0$. Data are from Ref.~\cite{PHENIXdAu06}. Theoretical 
results are calculated with different shadowing parameterizations (see text). 
}
\label{fig1}
\end{figure}

Figure \ref{fig1} displays recent PHENIX data 
(triangles with error bars)\cite{PHENIXdAu06} for
the most central \dAu\ collisions, where a high-$p_T$ suppression is
clearly seen. While incoherent multiscattering can only lead to enhancement
in $R_{dAu}$, nuclear shadowing displays two regions where an
$R_{dAu} < 1$ can be expected: (i) at small $x$ ($x \lesssim 0.2$), and 
(ii) in the EMC region ($0.5 \lesssim x \lesssim 0.9$)~\cite{emc95}. At 
RHIC energies the small-$x$ region is inconsequential at $p_T \gtrsim 6$~GeV/c. 
Thus we focus attention on the EMC effect as a possible mechanism for 
the measured suppression.  
Various shadowing parameterizations developed in the last 15 
years~\cite{Shad_HIJ,Shad_EKS,Shadxnw_uj,Shad_HKM,Shad_HKN} show 
different behaviors at small-$x$, but the EMC region appears  
rather robust in most models. 

To see the effect of the EMC region, we
calculate pion production in a wide momentum range. For this purpose we 
use a perturbative QCD improved parton model~\cite{Yi02}. The model is based 
on the factorization theorem and generates the invariant cross section as
a convolution of (nuclear) parton distribution functions $f_{a/A}$, 
perturbative QCD cross sections $\dd \sigma^{ab \rightarrow cd }/ \dd  \hat t$,  
and fragmentation functions $D_{\pi/c}$. We perform the 
calculation in leading order, following 
Refs.~\cite{Yi02,Bp02,Levai0306,Levai0611,bggqm04,bggqm05}:
\begin{eqnarray}
\label{hadX} 
E_{\pi} \frac{\dd \sigma_{\pi}^{dAu}}{\dd ^3p_{ \pi } } & = &
f_{a/d}(x_a,Q^2;{\bf k}_{Ta}) \otimes
f_{b/Au}(x_b,Q^2;{\bf k}_{Tb}) \otimes \nonumber \\
& \otimes & \frac{ \dd \sigma^{ab \rightarrow cd }}{\dd  \hat t }  
\otimes \frac{ D_{ \pi /c}(z_c,{\widehat Q}^2)}{\pi z_c^2} \,\, ,
\end{eqnarray}
where $Q^2$ and ${\widehat Q}^2$ represent the factorization and 
fragmentation scales, respectively, $x_a$, $x_b$, and $z_c$ are momentum fractions,
and ${\bf k}_T$-s stand for two-dimensional transverse momentum vectors. The
initial state effects of shadowing and multiscattering are included 
following the treatment in Refs.~\cite{Yi02,Levai0306,Levai0611}. 

Since the effects we investigate are on the $10-20 \%$ level, it is customary 
to present the obtained results on a linear scale in terms of the nuclear 
modification factor 
\begin{equation}
R_{dAu}(p_T)=\frac{1}{\langle N_{bin} \rangle } \cdot
\frac{E_{\pi} \dd \sigma^{dAu}_{\pi}/\dd^3 p_{\pi}}
{E_{\pi} \dd \sigma^{pp}_{\pi}/\dd^3 p_{\pi}}  \,\, .
\label{raa_def}
\end{equation}
Here $\langle N_{bin} \rangle$ is the average number of binary 
collisions in the various impact-parameter bins. 

\begin{figure}
\centerline{%
\rotatebox{0}{\includegraphics[width=\columnwidth,height=7.0truecm]%
   {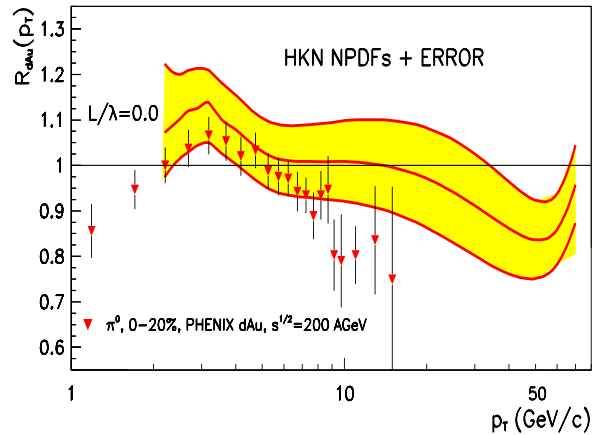}}
}
\caption{(Color online)
The influence of the uncertainty in the HKN
shadowing parameterization~\cite{Shad_HKN} on the factor $R_{dAu}$ in
the most central \dAu\ collisions.
Data are from Ref.~\cite{PHENIXdAu06}. 
}
\label{fig2}
\end{figure}
Together with the data in central \dAu\ collisions, we display 
our results with several shadowing parameterizations in Fig.\ref{fig1}.  
We use the HIJING shadowing including nuclear multiscattering (solid lines), 
the EKS shadowing (where multiscattering is represented by 
strong anti-shadowing) (dashed line), and the HKN parameterization 
(with and without nuclear multiscattering, dotted and dash-dotted 
lines, respectively). It can be seen in Fig.~\ref{fig1} that the 
suppression associated with the EMC effect shows up at transverse momenta
$p_T \gtrsim 20$~GeV/c in all models considered, and does not explain the
suppression in the data
at around $10$~GeV/c. The Cronin peak at $p_T \approx 3$~GeV/c is best 
reproduced by the HIJING parameterization. The ``HKN+multiscattering'' model
appears to overshoot the data at low $p_T$, while it gives very similar 
results to HIJING at $p_T \gtrsim 5$~GeV/c. Since the uncertainty of the 
HKN nuclear parton distribution functions is also available, we next 
examine the theoretical uncertainty of the description.     

In Fig.~\ref{fig2} the effect of the 
uncertainty given by the HKN shadowing parameterization is illustrated 
on the nuclear modification factor, $R_{dAu}$. The errors are calculated 
by the Hessian method, using the original code of the 
HKN group. The mean value as a function of $p_T$ is represented by a solid 
line, surrounded by an error band of approximately $\pm 10 \%$.
Although the data fall within this band at low $p_T$, for 
$p_T \gtrsim 8$~GeV/c the observed suppression is 
stronger than allowed by this calculation. 


\begin{figure}
\centerline{%
\rotatebox{0}{\includegraphics[width=\columnwidth,height=7.0truecm]%
   {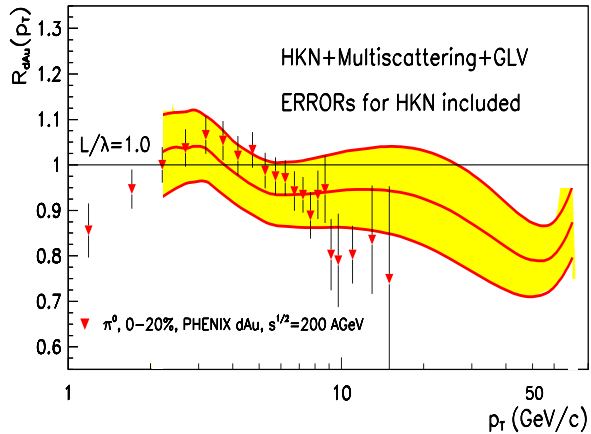}}
}
\caption{(Color online)
$R_{dAu}$ for ($0-20 \ \%$) \dAu\ collisions with jet energy loss
at opacity $L/\lambda _g= 1.0$.
Data are from Ref.~\cite{PHENIXdAu06}.
}
\label{fig3}
\end{figure}

Since we are not aware of any other initial-state 
suppression, we consider physics operating  
in the final state. In particular, jet energy loss suggests 
itself~\cite{glv1,glv2}. 
This idea is supported by the presence of a minor energy loss effect in 
peripheral \AuAu\ collisions~\cite{Barnafoldi:2006yv}. Assuming $L = 1.5$~fm 
for the average static transverse size of the traversed medium, and  
the usual GLV parameters of $\lambda = 1.5$~fm mean free path and 
$\mu = 0.5$~GeV screening mass, we calculate the effect of jet energy loss 
on the nuclear modification factor for central \dAu\ collisions. This is 
displayed in Fig.~\ref{fig3}. It can be seen that
inclusion of this energy loss results in a parallel down-shift of 
$R_{dAu}$ relative to curves in Fig~\ref{fig2}. The slope of 
the data as a function of $p_T$ is still very different from that of the 
calculated results. Nevertheless,
taking into account all experimental and theoretical uncertainties,
one could consider the displayed result with $L/\lambda = 1$ to be 
an acceptable compromise. 

Jet energy loss depends on the transverse parton density, which 
relates to the measurable hadronic quantity
$1/A_{\perp} \cdot \dd N_{ch}/ \dd y$, where $A_{\perp}$ is the transverse 
area of the deconfined region. Assuming a realistic geometry for central 
\AuAu\ and \dAu\ collisions and considering the experimental data on 
$\dd N_{ch}/ \dd y$, we obtain only a factor of $2$ difference for the
transverse parton densities between the \AuAu\ and \dAu\ cases. 
On this basis, one could expect the jet-quenching effect 
in \dAu\ to be even stronger than shown by the calculated band in Fig.~3.
However, jet energy loss in non-thermal matter is an open question,
which we plan to investigate in a forthcoming paper~\cite{GGB:2007}.

\begin{figure}
\centerline{%
\rotatebox{0}{\includegraphics[width=\columnwidth,height=7.0truecm]%
   {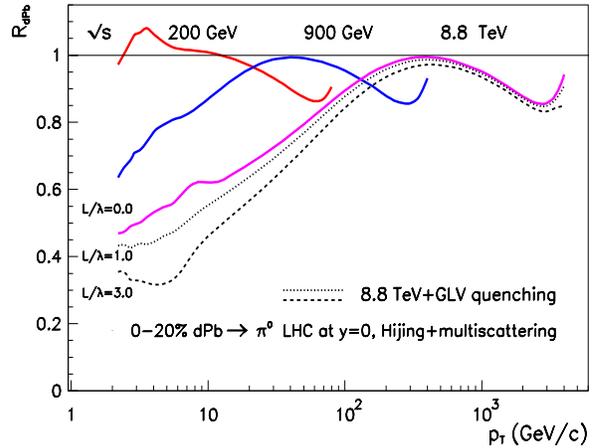}}
}
\caption{(Color online) The nuclear modification factor for pion production 
in $0-10\%$ most central $dPb$ collisions at different energies.
At $\sqrt{s_{NN}} = 8.8$ TeV the thin lines illustrate the effect of 
increasing energy loss.}
\label{fig4}
\end{figure}

\section{Predictions for the LHC}
\label{sec:2}

We expect that the EMC region will shift to higher and higher transverse 
momenta with increasing collision energy. At the same time, due to the
increasing parton density, the energy loss should also become larger.  
We repeat our calculation for $dPb$ collisions at $\sqrt{s_{NN}} = 900$~GeV 
and $8.8$~TeV to display these tendencies. The results are shown in Fig.~4.
Following the ordering of the lines on the left of the Figure, the
top curve represents the results with HIJING shadowing from 
Fig.~\ref{fig1}. The second line corresponds to 
a similar calculation (i.e. HIJING shadowing, no energy loss) at 
$900$~GeV c.m. energy. 
The next line shows the result at $8.8$~TeV 
without jet quenching, while the dotted and dashed lines 
illustrate the effect of energy loss. First we took 
$L = 1.5$ fm for the transverse size of the medium, and used the
value of $\lambda = 1.5$~fm as earlier. The introduction of a 
smaller $\lambda$ value to represent the increase in the transverse density 
of colored scattering centers with increasing c.m. energy gives
$L/\lambda = 3.0$. It can be seen that 
the dip corresponding to the EMC effect shifts to higher transverse momenta
with increasing energy as expected. While $R_{dAu}$ is above unity at 
$p_T \approx 3$ GeV/c for 0.2~TeV (Cronin effect), at $0.9$ and $8.8$~TeV,
where we are deep in the shadowing region at these transverse momenta,
at most a small ripple appears on $R_{dAu}$, which is rising towards
one with increasing transverse momentum in this region. The effect of varying 
$\lambda$ at $8.8$~TeV is minimal at transverse momenta $p_T \gtrsim 100$~GeV/c, 
but quite significant at $5$~GeV/c $\lesssim p_T \lesssim 10$~GeV/c.



\section{Conclusion}
\label{sec:3}

In conclusion, while this solution appears initially tempting, the EMC 
effect does not explain the unexpected suppression seen at 
$p_T \approx 10$~GeV/c in $R_{dAu}$ at $\sqrt{s_{NN}} = 200$~$A$GeV. Taking 
into account experimental uncertainties as well as the uncertainties of the 
HKN nuclear parton distribution functions, the experimental and theoretical 
results can be brought into agreement using a non-negligible amount of final 
state parton energy loss (with standard opacity parameters).
The non-thermal nature of the \dAu\ system casts some doubt on the parameter 
values. The presence of jet quenching in the \dAu\ system is somewhat 
surprising at first sight, but is justified by the 
similarity (within a factor of $2$) of the real transverse densities in 
the \AuAu\ and \dAu\ systems.
 

For LHC energies we do not have a fully reliable baseline calculation at 
present. It is clearly seen, however, that shadowing (i.e suppression) 
will dominate the momentum region up to $p_T \lesssim 100-200$~GeV/c. 
The effect of final state interactions (jet energy loss) may also reduce 
the nuclear modification factor, especially in this momentum region. The
EMC dip in the nuclear modification factor moves toward higher and higher 
transverse momenta with increasing energy.


\section*{Acknowledgments}

We thank Shunzo Kumano for providing  
the code and the data for calculating the error bars of the HKN 
parameterization of parton distribution functions.  
Our work was supported in part by Hungarian OTKA T043455, T047050, and NK62044, 
by the U.S. Department of Energy under grant U.S. DE-FG02-86ER40251,
and jointly by the U.S. and Hungary under MTA-NSF-OTKA OISE-0435701.


\end{document}